\documentclass[utf8]{FrontiersinHarvard} 

\usepackage{url,hyperref,lineno,microtype,subcaption}
\usepackage[onehalfspacing]{setspace}

\def\keyFont{\fontsize{8}{11}\helveticabold }
\def\firstAuthorLast{Weber {et~al.}} 
\def\Authors{Jana Weber\,$^{1}$, Marcel Weber\,$^{1}$, Juan Miguel Lopez Alcaraz\,$^{1,*}$}
\def\Address{$^{1}$AI4Health Division, Carl von Ossietzky Universität Oldenburg, Oldenburg, Germany}
\def\corrAuthor{Juan Miguel Lopez Alcaraz}
\def\corrEmail{juan.lopez.alcaraz@uol.de}

\usepackage{graphicx}
\usepackage{siunitx}
\usepackage{xcolor}
\usepackage{comment}

\usepackage{graphicx}
\usepackage{amsmath}
\usepackage{pifont}
\usepackage{amsmath, amssymb, amsfonts, amsthm}
\usepackage{graphicx}
\usepackage{amsfonts}       
\usepackage{nicefrac}       
\usepackage{microtype}      
\usepackage{mathtools}
\usepackage{comment}
\usepackage{pbox}
\usepackage{textcomp}
\usepackage[utf8]{inputenc}
\usepackage{multirow}
\usepackage{longtable}
\usepackage{geometry}
\usepackage{nameref}
\usepackage{textalpha}
\usepackage{booktabs} 
\usepackage{comment}
\usepackage{subfig}
\usepackage{threeparttable}
\usepackage{caption}

\usepackage{ragged2e} 

\usepackage[english]{babel}
\addto\extrasenglish{}

\begin{document}

\onecolumn
\firstpage{1}

\title[Depression Diagnosis from Patient Interviews]{Depression diagnosis from patient interviews using multimodal machine learning}

\author[\firstAuthorLast ]{\Authors} 
\address{} 
\correspondance{} 
\extraAuth{}

\maketitle

\begin{abstract}
\section{Background}
Depression is a major public health concern, affecting an estimated five percent of the global population. Early and accurate diagnosis is essential to initiate effective treatment, yet recognition remains challenging in many clinical contexts. Speech, language, and behavioral cues collected during patient interviews may provide objective markers that support clinical assessment.  
\section{Methods} 
We developed a diagnostic approach that integrates features derived from patient interviews, including speech patterns, linguistic characteristics, and structured clinical information. Separate models were trained for each modality and subsequently combined through multimodal fusion to reflect the complexity of real-world psychiatric assessment. Model validity was assessed with established performance metrics, and further evaluated using calibration and decision-analytic approaches to estimate potential clinical utility.  

\section{Results}
The multimodal model achieved superior diagnostic accuracy compared to single-modality models, with an AUROC of 0.88 and a macro F1-score of 0.75. Importantly, the fused model demonstrated good calibration and offered higher net clinical benefit compared to baseline strategies, highlighting its potential to assist clinicians in identifying patients with depression more reliably.  

\section{Conclusion} 
Multimodal analysis of patient interviews using machine learning may serve as a valuable adjunct to psychiatric evaluation. By combining speech, language, and clinical features, this approach provides a robust framework that could enhance early detection of depressive disorders and support evidence-based decision-making in mental healthcare.

\tiny
 \keyFont{ \section{Keywords:} depression diagnosis, digital biomarkers, multimodal analysis, machine learning, deep learning, clinical decision support}

\end{abstract}

\section{Introduction}

\subsection{Depression as a problem worldwide}
Depression represents a significant public health issue, impacting approximately 322 million individuals worldwide and accounting for 7.5\% of total years lived with disability \cite{world2017depression}. Untreated depression is associated with impaired quality of life, increased risk of comorbidities, and elevated mortality \cite{voros2020untreated}. Early and accurate diagnosis is essential to initiate effective treatment, yet recognition remains challenging in many clinical contexts due to subtle symptom presentation, variability across populations, clinical judgment, and commonly used self-report tools, such as which provide practical reference standards but are known to vary, particularly around diagnostic thresholds \cite{montano1994recognition}. Similarly, many diagnostic tools are based on hard binary thresholds, without detailing the level of the condition. Objective tools that can support clinicians in identifying depression have the potential to reduce diagnostic delays and improve treatment outcomes \cite{mao_2023}. Speech, language, and behavioral cues collected during patient interviews might represent promising sources of objective markers that may enhance clinical assessment \cite{smith2013diagnosis}.

\subsection{Machine learning in neuropsychiatry}
Machine learning (ML) has emerged as a transformative tool in neuropsychiatry, enabling analysis of complex, high-dimensional datasets to detect subtle patterns associated with psychiatric conditions for diagnosis \cite{strodthoff2024prospects} as well as for adverse event prediction in well-defined populations \cite{oloyede2024identifying}. ML models have been applied to various data modalities individually, but they particularly benefit from multimodal integration, supporting impactful predictive modeling in clinical settings such as emergency care \cite{alcaraz2025enhancing}. Advances in deep learning now allow more effective representation of longitudinal and multimodal data, capturing dependencies that are difficult to model with conventional statistical methods \cite{durstewitz2019deep}. In this context, early studies indicate that applying ML to behavioral and clinical data, such as speech and structured interviews, may provide actionable insights to assist clinicians in diagnosing depression more reliably \cite{li_2025}.

\subsection{Speech and text for depression detection}
Speech and language are rich sources of behavioral and cognitive information that can reflect an individual’s mental state. Audio features have been shown to capture important physiological and cognitive signals relevant for medical assessment \cite{henna2025interpretable}. In addition, lexical content and other linguistic characteristics correlate with depressive symptomatology \cite{losada2020evaluating}. Importantly, these data can be collected routinely during standard patient interviews, providing a non-invasive and cost-effective source of information \cite{gumus2023evaluating}. When combined with structured clinical and demographic data through multimodal approaches, such features can yield complementary insights and enhance the reliability of psychiatric application in clinical practice with depression as a significant task \cite{sui2023data}.

\subsection{Contributions}
In this work, we present a multimodal machine learning framework designed to support depression diagnosis during routine patient interviews. Our key contributions are: 1) We integrate speech, text, and structured clinical features to create a comprehensive representation of a patient’s mental state, leveraging data that can be collected non-invasively and without additional clinical burden. 2) We systematically evaluate both single-modality and multimodal models, assessing not only predictive performance but also calibration and potential clinical utility, ensuring the framework is informative for real-world decision-making. 3) We demonstrate that multimodal fusion enhances diagnostic accuracy over individual data sources, illustrating how combining complementary routine information can augment clinical assessment and support evidence-based psychiatric care.

\begin{figure}[ht!]
    \centering
    \includegraphics[width=\textwidth]{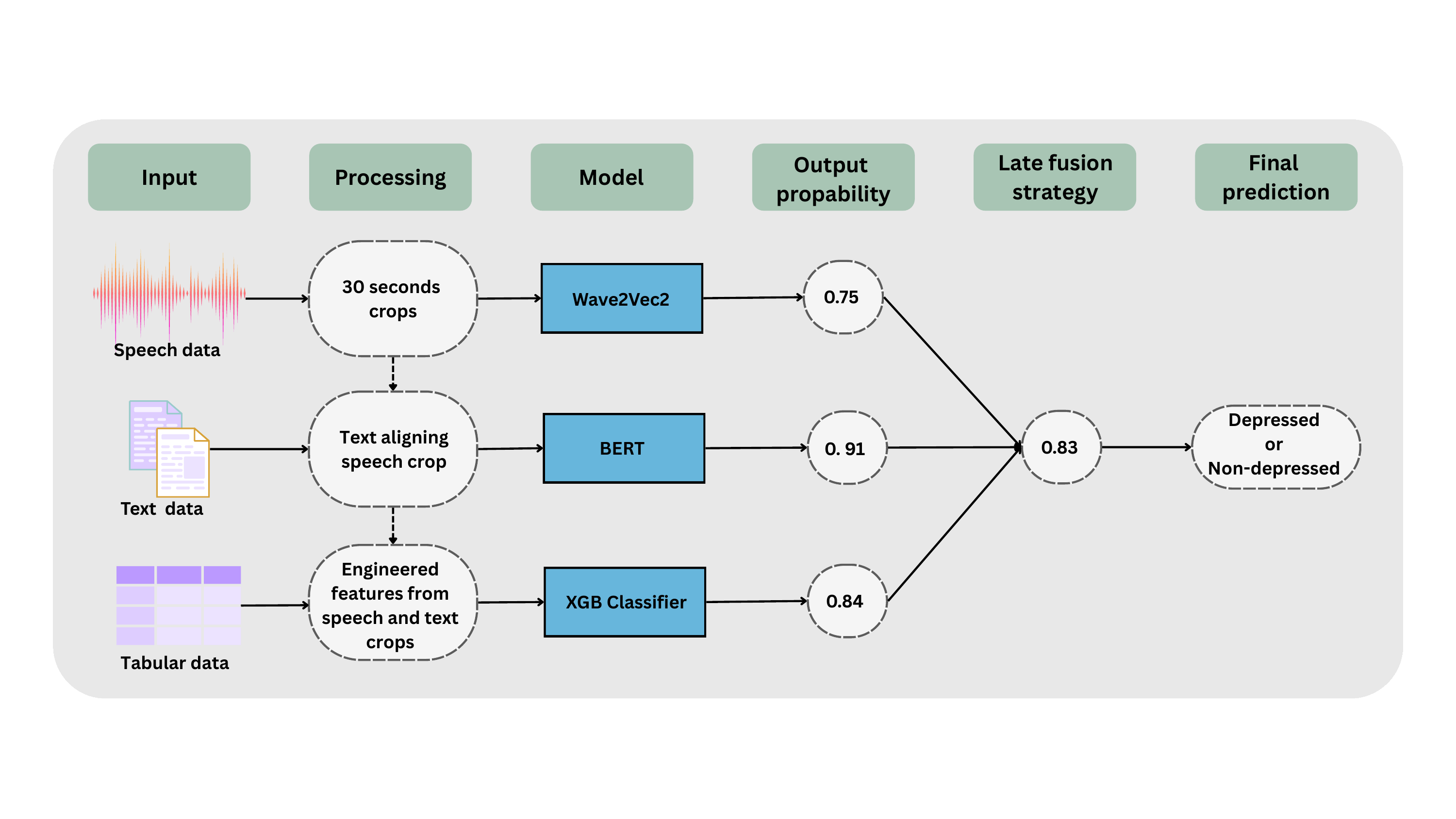}
    \caption{Conceptual overview of the implemented pipeline. The model integrates three modalities: audio, text, and tabular features. Preprocessing involves aligning 30-second segments and engineering tabular features from speech and text. Wave2Vec2, BERT, and XGBoost models each output class probabilities, which are then combined through late fusion to produce the final binary prediction of depression.}
    \label{fig:abstract}
\end{figure}

\section{Methods}

\subsection{Dataset}

\begin{table}[ht!]
\centering
\begin{tabular}{ll}
\toprule
\textbf{Variable} & \textbf{Total (n = 189)} \\
\midrule
\textbf{Demographics} \\
Female                     & 87 (46\%) \\
Male                       & 102 (54\%) \\
\midrule
\textbf{Audio} \\
Audio length, min [IQR]           & 15.92 [6.91-72.7] \\
\midrule
\textbf{Text} \\
Text length, characters [IQR]      & 14107.97 [5595-31505] \\
\midrule
\textbf{Tabular features} \\
Audio (Covarep) count      & 74 \\
Audio (Five formants) count    & 5 \\
Text count       & 7 \\
\midrule
\textbf{Label Prevalence} \\
Not depressed ($PHQ8 \leq 10$)                 & 133 (70\%) \\
Depressed ($PHQ8 > 10$)                 & 56 (30\%) \\
\bottomrule
\end{tabular}
\caption{Descriptive statistics of the study population ($n = 189$). Demographic variables are presented as counts and percentages. Audio and text characteristics are summarized as median [interquartile range (IQR)]. Tabular features report the number of extracted variables from audio and text modalities. Label prevalence is shown based on the PHQ-8 depression screening cutoff ($\leq 10$ = not depressed, $>10$ = depressed).}
\label{tab:descriptive_total}
\end{table}

Table \ref{tab:descriptive_total} contains descriptive statistics of the investigated dataset. We used the Distress Analysis Interview Corpus Wizard of Oz (DAIC-WOZ) dataset \cite{DAIC_one, DAIC_two}, developed to study verbal and nonverbal indicators of mental illness in structured interviews conducted by a virtual agent. The dataset includes 189 participants (102 males, 87 females), each with raw audio recordings (median length 15.9 min, IQR 6.9-72.7; total ~50 hours) and transcribed interviews (median 14,108 characters, IQR 5,595-31,505). Acoustic features were extracted using the Continuous VALence and REgression Platform (COVAREP) \cite{degottex2014covarep} and five formants, which capture vocal tract resonance frequencies \cite{boersma2001praat}. Depression severity was assessed via the PHQ-8 (Patient Health Questionnaire-8) \cite{kroenke2009phq}, with 30\% as depressed and 70\% as controls. We use a train, validation, and test set splits of 107, 35, and 47 patient interviews respectively. Overall, the dataset provides rich multimodal information that can be leveraged for machine learning-based depression detection using routine interview data without additional invasive assessments.

\subsection{Features}
We grouped features into three categories: raw audio (16 kHz unprocessed speech), raw text (transcribed speech), and tabular features derived from audio (COVAREP and formants) and text (lexical metrics), totaling 550 features (543 audio, 7 text). For model training, we randomly cropped 30-second segments per interview, aligning corresponding text and tabular features, which allowed multiple samples per visit and improved model robustness. Audio features were extracted at 10 ms intervals and aggregated over the crop. To evaluate modality contributions, we tested seven configurations: (1) audio only, (2) text only, (3) tabular only, (4) audio + text, (5) audio + tabular, (6) text + tabular, and (7) full multimodal fusion. This setup enables systematic comparison of unimodal versus multimodal approaches. Full feature descriptions are provided in the supplementary material.

\subsection{Target}
The dataset is labeled using the PHQ-8 score, which is based on participants' responses to eight questions assessing symptoms such as mood and appetite \cite{kroenke2009phq}. For analysis, a binary label was created using a threshold of 10: participants with a score above 10 were classified as exhibiting depressive symptoms, while those with a score of 10 or lower were considered non-depressed. This threshold is widely used in clinical and research settings to identify individuals at risk for depression, enabling straightforward binary classification in subsequent modeling.

\subsection{Models}
For the audio model, speech waveforms were processed at 16 kHz in 30-second chunks with 50\% overlap. We used a pretrained Wav2Vec2 model, and each chunk produced a binary prediction ("depressed" or "not depressed"). Final patient-level predictions were computed by averaging the chunk-level outputs. For the text model, transcripts were concatenated into a single sequence per patient, tokenized with a maximum length of 256 tokens, and processed using a BERT model. The output was binary, similar to the audio model. Tabular features included both audio-derived (COVAREP and formants) and text-derived features (extracted via SpaCy). These were combined into a single table per patient and modeled using XGBoost with cross-validation for evaluation. For multimodal modeling, we applied late fusion by computing a weighted average of the outputs from each modality, followed by calibration using logistic regression. This approach allowed systematic integration of complementary information from audio, text, and tabular features. Further model hyperparameter configurations are provided in the supplementary material.

\subsection{Performance evaluation}
We assessed model performance primarily using the area under the receiver operating characteristic curve (AUROC), a widely adopted ranking-based metric that is robust to class imbalance. Specifically, we report the macro AUROC, which captures overall discriminative performance without requiring predefined decision thresholds. Recent studies \citep{mcdermott2024closer} have highlighted AUROC’s advantages over alternatives such as the area under the precision-recall curve (AUPRC), particularly in imbalanced settings. Confidence intervals (95\%) were estimated via bootstrapping on the test set. Complementary metrics, including precision, recall, and F1-score, were also reported to provide a nuanced assessment of predictive performance. Precision measures the proportion of true positives among all positive predictions, recall quantifies the ability to detect all actual positive cases, and the F1-score, as the harmonic mean of precision and recall, is particularly informative in class-imbalanced data. These metrics are critical in our context, where false negatives, missed cases of depression, carry important clinical consequences.

We further evaluated model calibration using calibration curves to assess the agreement between predicted probabilities and observed outcomes. Clinical utility was quantified via decision curve analysis (net benefit), comparing model performance to two baseline strategies: referring all patients versus referring none. As age information was unavailable, we focused on gender as the primary demographic feature. Following previous work on fairness in machine learning \cite{pessach2022review}, we included demographic parity by gender through a stratified analysis using the distinctive acoustic feature, fundamental frequency (F0) where we report Equal Opportunity and Equalized Odds metrics, based on the true positive rate (TPr, the proportion of actual positive cases correctly identified) and the false positive rate (FPr, the proportion of actual negative cases incorrectly classified as positive) for each gender. These measures provide complementary insights into how fairly the model performs across genders.

\section{Results}

\subsection{Discriminative performance}

\begin{table}[ht!]
\centering
\sisetup{detect-all}
\NewDocumentCommand{\B}{}{\fontseries{b}\selectfont}
\begin{tabular}{
  @{}
  l
  S[table-format=1.2]
  S[table-format=1.2]
  S[table-format=1.2]
  S[table-format=1.2]
  S[table-format=-1.2]
  S[table-format=1.2]
  S[table-format=1.2]
  S[table-format=1.2]
  S[table-format=1.2]
  S[table-format=1.2]
  @{}
}
& \multicolumn{8}{c}{Metrics} \\
\cmidrule(l){2-11}
& \multicolumn{2}{c}{Precision} & \multicolumn{2}{c}{Recall} & \multicolumn{2}{c}{F1 Score} & \multicolumn{2}{c}{Macro F1} & \multicolumn{2}{c}{AUROC}\\
\cmidrule(l){2-7}
 & {C} & {D} & {C} & {D} & {C} & {D} \\
\midrule
Audio & 0.66 & 0.21 & 0.66 & 0.21 & 0.66 & 0.21 & \multicolumn{2}{c}{0.44} & \multicolumn{2}{c}{0.42} \\
Text & 0.76 & 0.41 & 0.69 & 0.50 & 0.72 & 0.45 & \multicolumn{2}{c}{0.59} & \multicolumn{2}{c}{0.60} \\
Tabular & 0.77 & 0.40 & 0.62 & 0.57 & 0.69 & 0.47 & \multicolumn{2}{c}{0.58} & \multicolumn{2}{c}{0.57} \\
\midrule
Audio / Text & 0.82 & 0.67 & 0.88 & 0.57 & 0.85 & 0.62 & \multicolumn{2}{c}{0.73} & \multicolumn{2}{c}{0.73}\\
Audio / Tabular & 0.81 & 0.70 & 0.91 & 0.50 & 0.85 & 0.58 & \multicolumn{2}{c}{0.72} & \multicolumn{2}{c}{0.84} \\
Text / Tabular & 0.79 & 0.54 & 0.81 & 0.50 & 0.80 & 0.52 & \multicolumn{2}{c}{0.66} & \multicolumn{2}{c}{0.72} \\
\midrule
Audio / Text / Tabular & \B0.83 & \B0.73 & \B0.91 & \B0.57 & \B0.87 & \B0.64 & \multicolumn{2}{c}{\B0.75} & \multicolumn{2}{c}{\B0.88} \\
\end{tabular}
\captionsetup{skip=10pt} 
\caption{Discriminative performance of models across different modalities. Results are reported separately for control (C) and depressed (D) classes in terms of precision, recall, and F1 score, alongside macro-F1 and AUROC as global metrics. Single-modality models (Audio, Text, Tabular) are compared with multimodal combinations. Overall, multimodal approaches outperform unimodal ones, with the integration of Audio, Text, and Tabular features achieving the best performance across most metrics.}

\label{tab:discriminative}
\end{table}

Table \ref{tab:discriminative} summarizes the discriminative results. Among unimodal models, text features performed best (F1 = 0.59, AUROC = 0.60), followed closely by tabular features (F1 = 0.47, AUROC = 0.57). The audio-only model showed the weakest performance (F1 = 0.44, AUROC = 0.42), mainly due to very low precision for the depressed class. Multimodal approaches consistently outperformed unimodal ones. The best results were obtained by combining all three modalities, reaching an AUROC of 0.88 and a macro F1 score of 0.75. Pairwise combinations also improved performance, with audio+tabular slightly stronger (AUROC = 0.84) than audio+text or text+tabular. Overall, multimodal fusion clearly enhanced both discrimination and balance across classes, mitigating the weaknesses of single modalities.

\begin{figure}[ht!]
    \centering
    \includegraphics[width=0.5\linewidth]{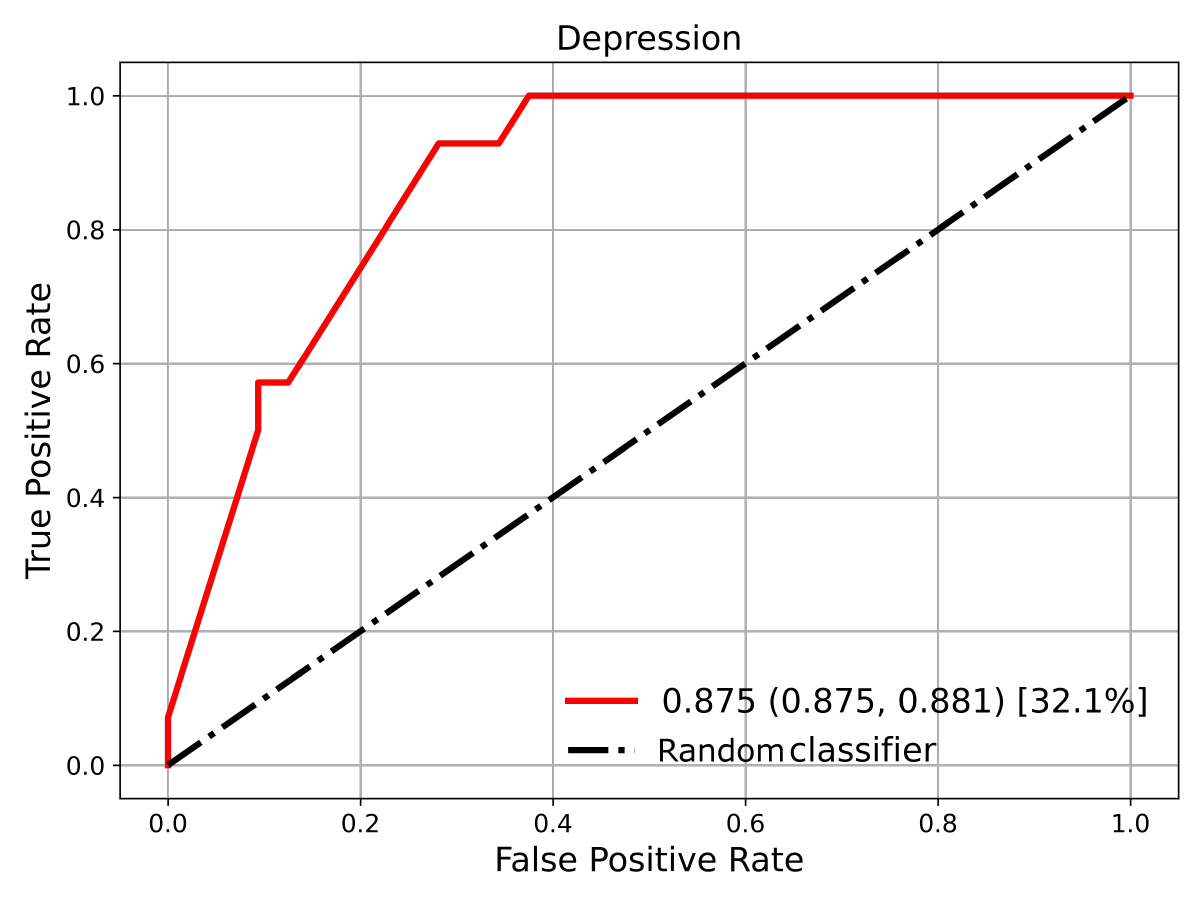}
    \caption{Receiver operating characteristic (ROC) analysis. The red solid line shows the mean AUROC with 95\% confidence intervals estimated via empirical bootstrap resampling. The black dashed line represents the performance of a random classifier (AUROC = 0.5) as a reference.}
    \label{fig:auroc}
\end{figure}

Multimodal integration consistently improved performance compared to unimodal models, where the best results were achieved by combining all three modalities, with an AUROC of 0.88 (Figure \ref{fig:auroc}). This model showed strong detection of non-depressed individuals (F1 = 0.87) and moderate but improved detection of depressed cases (F1 = 0.64). Overall, multimodal fusion mitigates the weaknesses of unimodal inputs, though performance remains better for the non-depressed class. While this class imbalance is a limitation, reliable identification of non-depressed individuals can still be clinically useful, as it helps reduce unnecessary referrals and ensures resources are focused on higher-risk cases.

\begin{table}[ht!]
\centering
\begin{tabular}{lcc}
\toprule
\textbf{PHQ-8 threshold} & \textbf{Performance} & \textbf{Baseline} \\
\midrule
$\geq 5$  & 0.79 AUROC~($\uparrow$) & 0.50 (Random) \\
$\geq 10$ & 0.88 AUROC~($\uparrow$) & 0.50 (Random) \\
$\geq 15$ & 0.80 AUROC~($\uparrow$) & 0.50 (Random) \\
\midrule
Regression & 5.381 MAE~($\downarrow$) & 6.403 (Train SD) \\
\bottomrule
\end{tabular}
\caption{Model performance in classification (AUROC, higher is better) across PHQ-8 thresholds and regression (MAE, lower is better) compared to baseline values.}
\label{tab:threshold_regression_analysis}
\end{table}

Table \ref{tab:threshold_regression_analysis} shows the model’s performance in classification and regression tasks. For classification, AUROC values were 0.79, 0.88, and 0.80 at PHQ-8 thresholds of $\geq$5, $\geq$10, and $\geq$15, all exceeding the random baseline (0.50) where the standard threshold used in current practice $\geq$10 achieve the best performance. Regression analysis yielded an MAE of 5.381, lower than the training set’s standard deviation (6.403) and mean (5.468), indicating reliable estimation of continuous PHQ-8 scores.

\subsection{Stratified analysis}

\begin{figure}[ht!]
    \centering
    \includegraphics[width=0.5\linewidth]{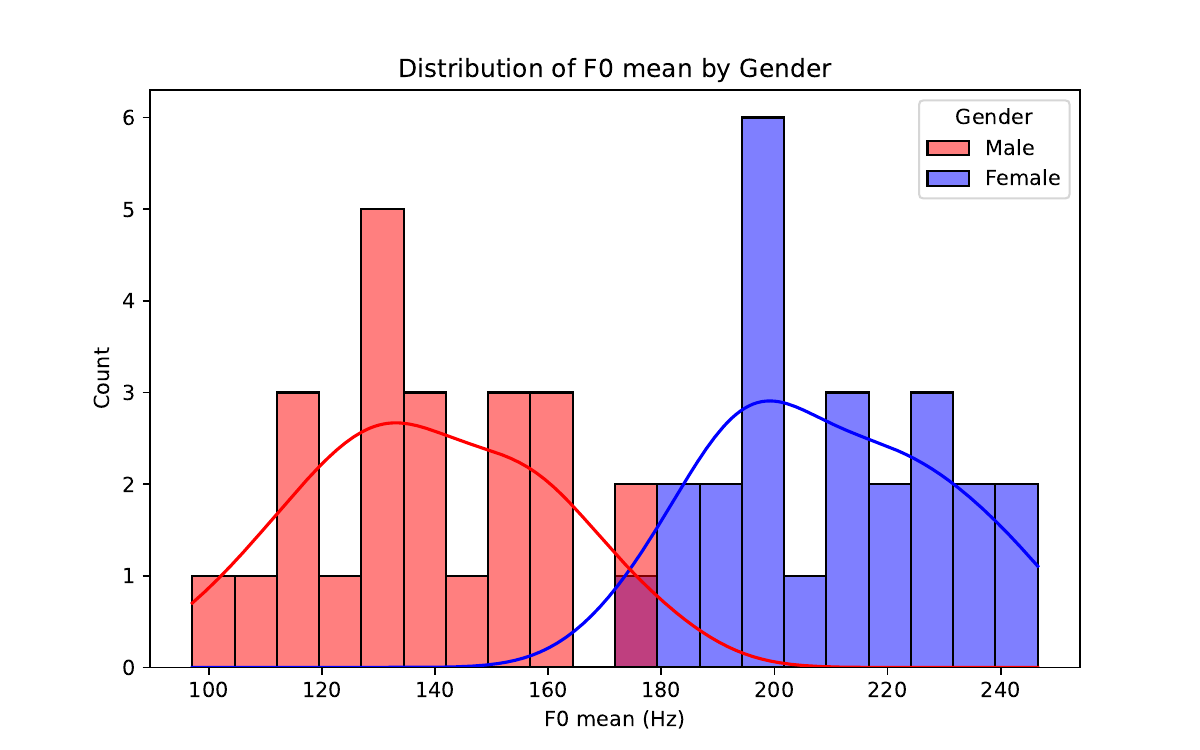}
    \caption{Distribution of the fundamental frequency (F0, i.e., pitch) across genders. Male voices (red) show a lower range (100–180 Hz), while female voices (blue) are generally higher (180–240 Hz).}
    \label{fig:f0_gender}
\end{figure}

Figure \ref{fig:f0_gender} shows the distribution of F0 by gender, with male voices concentrated between 100–180 Hz and female voices between 180–240 Hz, highlighting F0’s discriminative power as a gender-related acoustic feature. Evaluating model performance separately by gender reveals notable differences: females show perfect precision (1.0) but low recall (0.14) and AUROC 0.57, whereas males achieve high recall (1.0), precision 0.70, and AUROC 0.91. Although the model uses all features, this performance gap aligns with F0 differences—males’ lower baseline and wider spread provide stronger cues, while the narrower, higher-range female distribution limits contribution. Following \cite{pessach2022review}, we computed true positive ratio (TPr) and true negative ratio (TNr) to assess fairness: females have TPr 0.14 and TNr 1.0, while males have TPr 1.0 and TNr 0.81. This indicates that the model is more likely to correctly identify positive male cases, while negative female cases are more accurately classified, reflecting a subgroup performance imbalance rather than equal treatment.

\subsection{Calibration}

\begin{figure}[ht!]
    \centering
    \includegraphics[width=0.5\linewidth]{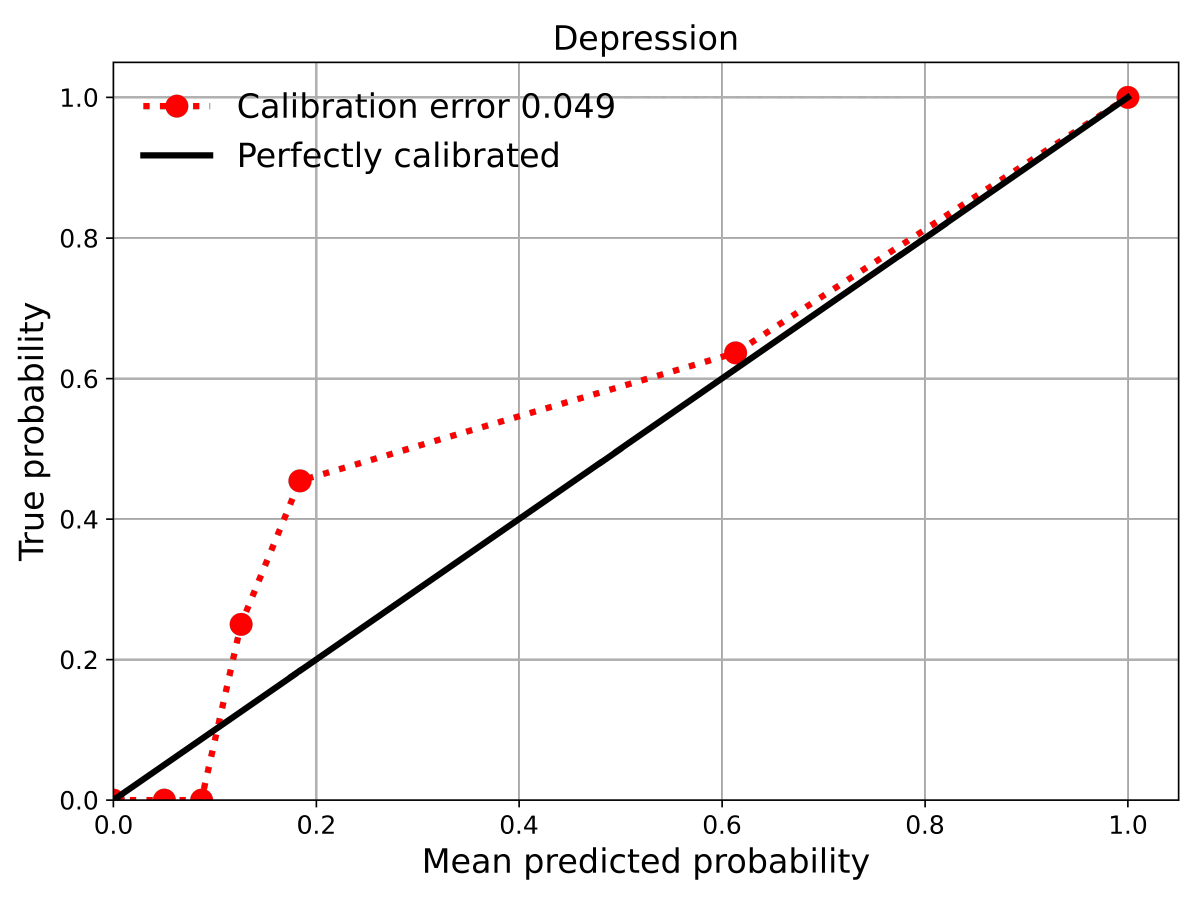}
    \caption{Calibration curve. The red solid line depicts the observed calibration performance of the model, while the black solid line represents a perfectly calibrated classifier (ideal reference). Overall, the model shows good calibration, with predictions in the higher probability range aligning almost perfectly with the true outcomes. This suggests particularly reliable performance for the positive (depressed) class in the upper probability range.}
    \label{fig:calibration}
\end{figure}

Figure \ref{fig:calibration} shows the calibration curve of our model, crucial for medical applications. Overall, the model shows good calibration. Initially, some points are above the diagonal, which means that the model is underconfident and underestimates risks and probabilities at these points. In general, with a calibration error of 0.04, it can be said that the model is fundamentally good, as the average difference between the predicted probabilities and the actual frequency is 4\%. The model can be trusted, but there is still room for improvement before it can be deployed in practice.

\subsection{Decision analysis}

\begin{figure}[ht!]
    \centering
    \includegraphics[width=0.5\linewidth]{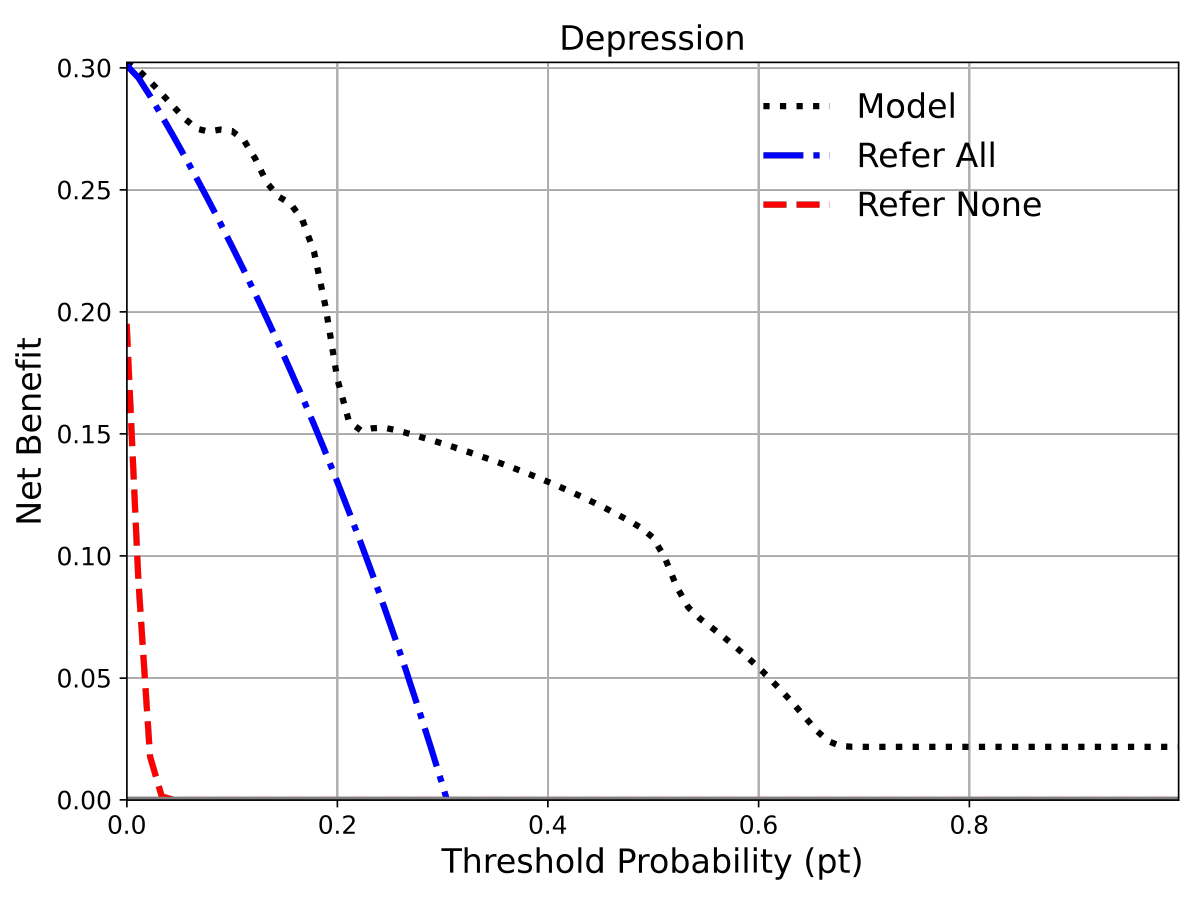}
    \caption{Decision curve analysis. The black dashed line shows the net benefit of the proposed model across a range of threshold probabilities. The blue dashed line represents the strategy of referring all individuals, while the red dashed line represents the strategy of referring none. The model achieves higher net benefit than both reference strategies across the entire range, indicating superior clinical usefulness.}
    \label{fig:decision_analysis}
\end{figure}

Figure \ref{fig:decision_analysis} presents the decision curve (net benefit analysis), which complements traditional performance measures such as the AUROC by incorporating the clinical consequences of different decision strategies. The black dashed line shows the net benefit of our model, while the blue and red dashed lines correspond to the strategies of treating all or treating none, respectively. Across a wide range of threshold probabilities, the model provides greater net benefit than either reference strategy, indicating its potential clinical usefulness. In particular, the “treat all” strategy (blue line) declines sharply, underscoring the potential harm of unnecessary treatment. The model achieves its highest net benefit at low threshold probabilities, suggesting that it may be especially valuable for identifying patients at risk of depression early, when a lower decision threshold for intervention is clinically appropriate.

\section{Discussion}

\subsection{Methodological findings}
Our results highlight the importance of multimodality in healthcare applications, particularly for psychological diagnosis. As shown in Table \ref{tab:discriminative}, combining modalities substantially outperforms unimodal models. The full multimodal setup improved the macro F1 score by 0.31 and the AUROC by 0.44 compared to the weakest unimodal baseline. Comparisons with prior work on the same dataset further emphasize the role of methodological rigor. \cite{burdisso2024} reported an F1-score of 0.90 but relied on validation set results rather than a held-out test set, inflating performance estimates. Similarly, \cite{dai_improving_2021} achieved an F1-score of 0.96 on the development set using a multimodal pipeline with audio, video, and semantic features. However, their score dropped to 0.67 on the test set, suggesting overfitting and limited generalization. In contrast, our models show more stable performance despite relying on even fewer modalities. Among unimodal models, raw audio alone was less predictive than text or tabular features, likely because acoustic markers of depression (e.g., pitch, prosody) are subtle and variable across speakers. For instance, research has shown that depressed individuals often exhibit lower pitch variability and slower speech rates, which can be subtle and variable across speakers, making them challenging to capture consistently in raw audio alone \cite{di2024unraveling}. However, when combined with other modalities, audio consistently enhanced performance and contributed significant complementary information. Text and tabular features, being more structured and explicit, provided stronger standalone discriminative power, but it was the integration of all three modalities that yielded the most robust and generalizable outcomes. This aligns with recent work in multimodal representation learning \cite{yang2025fine,yang2024enhancing}, which highlights the importance of structured multimodal fusion for complex psychological states. Similarly, advances in cross-modal feature learning \cite{cui2024enhancing,yang2024cross} emphasize that complementary information across modalities is most beneficial when representations are aligned and robustly integrated.

\subsection{Clinical findings}
Our results demonstrate that significant diagnostic signals of depression can be extracted from routine data such as audio recordings, clinical text, and structured patient information. This finding has several important implications for clinical practice. First, multimodal AI models can support faster and lower-cost screening, reducing the reliance on extensive manual assessments \cite{Khanna2022}. By leveraging routinely collected data, such systems could provide early warnings during regular consultations or remote interactions, enabling earlier diagnosis and intervention. Second, such tools can help mitigate diagnostic bias. Unlike a single clinician’s evaluation, the models learn from patterns across a wide population of patients and evaluators, offering a more standardized and less subjective perspective. This could be especially valuable in settings where access to specialized mental health professionals is limited \cite{saxena2007resources}. Finally, integration into routine practice can be envisioned in specific use cases: for example, as a decision support tool in primary care to flag at-risk patients for referral, as an adjunct in telemedicine platforms to enhance remote consultations, or as part of longitudinal monitoring systems that track patients’ risk levels over time \cite{rony_artificial_2025}. In all cases, the goal is not to replace clinical judgment but to provide an additional, reliable source of evidence that enhances the timeliness and equity of mental health care.

\subsection{Limitations}

This study has two main limitations. First, the sample size remains relatively modest, which restricts the robustness of the findings and may limit the ability to fully capture the heterogeneity of depressive symptoms across large populations despite our augmentation sampling approach. Larger datasets are needed to confirm the stability of the reported performance gains \cite{collins2016sample}. Second, the models have not yet undergone external validation on independent cohorts. Without such testing, the generalizability of the results to different clinical settings, languages, or patient demographics remains uncertain. Future work should therefore emphasize replication across larger and more varied populations to ensure clinical applicability \cite{riley2024evaluation}. Third, our analysis indicates that the model performs less effectively for female patients, particularly in identifying positive cases, highlighting a potential gender bias. Addressing this limitation will require incorporating additional features or strategies to improve fairness and ensure equitable model performance across genders.

\subsection{Future work}

Several avenues can be pursued to extend this work. First, incorporating additional modalities, such as physiological signals, could further improve predictive performance and provide complementary information. Second, enhancing explainability is a key goal. This includes analyzing patients classified as non-depressed or those with co-occurring conditions \cite{ott2024using}, as well as aiming for causal attributions rather than purely associational insights. Recent work on causal explanations in time series \cite{alcaraz2024causalconceptts} could be adapted to multimodal speech-based frameworks, although a new model design would be required. Third, moving beyond binary classification, future models could predict graded levels of depression, enabling risk stratification and more nuanced clinical decision-making for situations like treatment administration \cite{duval2006treatments}. Fourth, integrating additional demographic and clinical variables would allow for patient-specific predictions, supporting personalized mental health care. Finally, exploring alternative PHQ-8 thresholds for defining binary outcomes could inform how varying diagnostic criteria influence model performance and practical applicability. In our current approach, each modality was modeled independently, and alignment between text and speech was approximated at the level of interview crops. This inevitably leaves portions of audio and text that may not correspond directly, potentially limiting the effectiveness of multimodal fusion. Future work could leverage semantic alignment methods \cite{yang2023semantic,yang2022implicit} to explicitly map audio, text, and structured features into a shared latent space. Such techniques would ensure that information from different modalities is synchronized at a finer-grained semantic level, improving both robustness and interpretability of the predictions.

\section{Conclusion}

We demonstrate that multimodal analysis of routine patient interviews which combines audio, text, and structured clinical data can effectively support depression detection. Multimodal fusion outperforms single modalities, enabling faster, low-cost, and objective screening while reducing reliance on a single clinician’s judgment. These results highlight the potential of ML-driven tools to enhance early diagnosis, support clinical decision-making, and improve the way for personalized mental health care.

\section*{Conflict of Interest Statement}
Upon manuscript submission, all authors completed the author disclosure form, confirming the absence of any conflicts of interest.

\section*{Author Contributions}
J.W.: Writing - original draft, Conceptualization, Methodology, Investigation, Data curation, Software, Validation, Visualization; M.W.: Writing - original draft, Conceptualization, Methodology, Investigation, Data curation, Software, Validation, Visualization; J.M.L.A.: Writing - review \& editing, Conceptualization, Methodology, Supervision, Project administration, Resources.

\section*{Funding}
This research did not receive any specific grant from funding agencies in the public, commercial, or not-for-profit sectors.

\section*{Supplemental Data}
 \href{http://home.frontiersin.org/about/author-guidelines#SupplementaryMaterial}{Supplementary Material} should be uploaded separately on submission, if there are Supplementary Figures, please include the caption in the same file as the figure. LaTeX Supplementary Material templates can be found in the Frontiers LaTeX folder.

\section*{Data Availability Statement}
The datasets analyzed for this study can be found in the DAIC-WOZ repository \url{https://dcapswoz.ict.usc.edu/}. Code for dataset preprocessing and experimental replications can be found on dedicate code repository \url{https://github.com/UOLMDA2025/Depression}.

\bibliographystyle{Frontiers-Harvard} 
\bibliography{test}

\clearpage

\appendix

\section{Supplementary Tables and Figures}

\subsection{Tables}

\begin{longtable}{l|p{8cm}}
\caption{Overview of Covarep acoustic features.} \label{tab:features_covarep} \\
\hline
\textbf{Feature Name} & \textbf{Description} \\
\hline
\endfirsthead

\hline
\textbf{Feature Name} & \textbf{Description} \\
\hline
\endhead

\hline
\multicolumn{2}{r}{\emph{Continued on next page}} \\
\endfoot

\hline
\endlastfoot

F0 & Fundamental frequency (pitch) of the voice. \\
VUV & Voiced/unvoiced decision. \\
NAQ & Normalized Amplitude Quotient, a glottal flow measure. \\
QOQ & Quasi-Open Quotient, related to glottal opening. \\
H1H2 & Difference in amplitude between the first and second harmonics. \\
PSP & Peak Slope Parameter, reflects harmonic richness. \\
MDQ & Maxima Dispersion Quotient, a measure of glottal asymmetry. \\
peakSlope & Slope of the harmonic peaks. \\
Rd & Glottal shape parameter. \\
Rd\_conf & Confidence measure for the Rd parameter. \\
creak & Creaky voice probability. \\
MCEP\_{0-24} & Mel cepstral coefficient {0-24} (spectral envelope). \\
HMPDM\_{0-24} & Harmonic model phase distortion mean {0-24}. \\
HMPDD\_{0-12} & Harmonic model phase distortion deviation {0-12}. \\
\hline
\end{longtable}

\begin{longtable}{l|p{8cm}}
\caption{Overview of formant features.} \label{tab:features_formants} \\
\hline
\textbf{Feature Name} & \textbf{Description} \\
\hline
\endfirsthead

\hline
\textbf{Feature Name} & \textbf{Description} \\
\hline
\endhead

\hline
\endfoot

\hline
\endlastfoot

F1 & First formant – lowest vocal tract resonance frequency. \\
F2 & Second formant – tongue position and vowel quality. \\
F3 & Third formant – vocal tract shape. \\
F4 & Fourth formant – high-frequency resonance. \\
F5 & Fifth formant – additional vocal tract characteristics. \\
\hline
\end{longtable}

\begin{longtable}{l|p{8cm}}
\caption{Overview of text-based features.} \label{tab:features_text} \\
\hline
\textbf{Feature Name} & \textbf{Description} \\
\hline
\endfirsthead

\hline
\textbf{Feature Name} & \textbf{Description} \\
\hline
\endhead

\hline
\endfoot

\hline
\endlastfoot

TTR & The type-token-ratio, the ratio of unique words (types) to the total number of words (tokens) in a text. \\
avg\_sentence\_length & The average number of words per sentence. \\
past\_tense\_ratio & The proportion of verbs in past tense relative to all verbs in the text. \\
pronoun\_count & The total number of personal pronouns in the text. \\
mean\_local\_coherence & The average semantic similarity between adjacent sentences or clauses (often computed using embeddings). \\
filler\_word\_count & The number of non-content words used to fill pauses (e.g., "um", "uh", "like", "you know"). \\
sentiment & A numeric or categorical measure of emotional tone (e.g., positive, negative, neutral), often derived using sentiment analysis tools. \\
\hline
\end{longtable}

\end{document}